\documentclass[aps,prl,reprint,preprintnumbers,superscriptaddress,amsmath,amssymb,bibnotes,longbibliography]{revtex4-2}
\usepackage{graphicx}
\usepackage{dcolumn}
\usepackage{bm}
\usepackage[colorlinks,linkcolor=blue,anchorcolor=blue,citecolor=blue,breaklinks,CJKbookmarks=True,urlcolor=blue,filecolor=blue,menucolor=blue,runcolor=blue]{hyperref}

\newcommand{\tcr}[1]{\textcolor{black}{#1}}

\begin{document}

\preprint{Preprint: \today}

\title{NbReSi: A Noncentrosymetric Superconductor with Large Upper Critical Field}

\author{H. Su}
\affiliation  {Center for Correlated Matter and Department of Physics, Zhejiang University, Hangzhou 310058, China}
\affiliation  {Zhejiang Province Key Laboratory of Quantum Technology and Device, Department of Physics, Zhejiang University, Hangzhou 310058, China}
\author{T. Shang}
\email[Corresponding author: ]{tshang@phy.ecnu.edu.cn}
\affiliation  {Key Laboratory of Polar Materials and Devices (MOE), School of Physics and Electronic Science, East China Normal University, Shanghai 200241, China}
\author{F. Du}
\affiliation  {Center for Correlated Matter and Department of Physics, Zhejiang University, Hangzhou 310058, China}
\affiliation  {Zhejiang Province Key Laboratory of Quantum Technology and Device, Department of Physics, Zhejiang University, Hangzhou 310058, China}
\author{C. F. Chen}
\affiliation  {Center for Correlated Matter and Department of Physics, Zhejiang University, Hangzhou 310058, China}
\affiliation  {Zhejiang Province Key Laboratory of Quantum Technology and Device, Department of Physics, Zhejiang University, Hangzhou 310058, China}
\author{H. Q. Ye}
\affiliation  {Center for Correlated Matter and Department of Physics, Zhejiang University, Hangzhou 310058, China}
\affiliation  {Zhejiang Province Key Laboratory of Quantum Technology and Device, Department of Physics, Zhejiang University, Hangzhou 310058, China}
\author{X. Lu}
\affiliation  {Center for Correlated Matter and Department of Physics, Zhejiang University, Hangzhou 310058, China}
\affiliation  {Zhejiang Province Key Laboratory of Quantum Technology and Device, Department of Physics, Zhejiang University, Hangzhou 310058, China}
\affiliation  {Collaborative Innovation Center of Advanced Microstructures, Nanjing University, Nanjing, 210093, China}
\author{C. Cao}
\affiliation  {Center for Correlated Matter and Department of Physics, Zhejiang University, Hangzhou 310058, China}
\affiliation  {Condensed Matter Group, Department of Physics, Hangzhou Normal University, Hangzhou 311121, China}
\author{M. Smidman}
\affiliation  {Center for Correlated Matter and Department of Physics, Zhejiang University, Hangzhou 310058, China}
\affiliation  {Zhejiang Province Key Laboratory of Quantum Technology and Device, Department of Physics, Zhejiang University, Hangzhou 310058, China}
\author{H. Q. Yuan}
\email[Corresponding author: ]{hqyuan@zju.edu.cn}
\affiliation  {Center for Correlated Matter and Department of Physics, Zhejiang University, Hangzhou 310058, China}
\affiliation  {Zhejiang Province Key Laboratory of Quantum Technology and Device, Department of Physics, Zhejiang University, Hangzhou 310058, China}
\affiliation  {Collaborative Innovation Center of Advanced Microstructures, Nanjing University, Nanjing, 210093, China}
\affiliation  {State Key Laboratory of Silicon Materials, Zhejiang University, Hangzhou 310058, China}

\date{\today}



\begin{abstract}
We report the discovery of superconductivity in noncentrosymmetric NbReSi, which crystallizes in a hexagonal ZrNiAl-type crystal structure with space group $P\bar{6}2m$ (No.~189). Bulk superconductivity, with $T_c$ = 6.5\,K was characterized via electrical-resistivity, magnetization, and heat-capacity measurements.
The low-temperature electronic specific heat suggests a fully gapped superconducting state in NbReSi, while a large upper critical field of $\mu_0H_\mathrm{c2}(0)$ $\sim$ 12.6\,T is obtained, which is comparable to the weak-coupling Pauli limit. The electronic band-structure calculations show that the density of states at the Fermi level are dominated by Re and Nb $d$-orbitals, with a sizeable band splitting induced by the antisymmetric spin-orbit coupling.
NbReSi represents another candidate material for revealing the puzzle of time-reversal symmetry breaking observed in some Re-based superconductors and its relation to the lack of inversion symmetry.

\end{abstract}

\maketitle

\section{Introduction\label{sec:introduction}}\enlargethispage{8pt}

Noncentrosymmetric superconductors (NCSC), where the crystal structure lacks an inversion center, have been widely investigated after the discovery of superconductivity (SC) in heavy fermion compound CePt$_3$Si~\cite{BauerPhysRevLett2004}.
The lack of inversion symmetry gives rise to antisymmetric spin orbit coupling (ASOC), which causes band splitting near the Fermi level. As a consequence, the superconducting pairing may be a mixture of spin singlet and spin triplet states~\cite{bauer2012non,Smidman2017}. 
Unconventional physical properties closely related to such a mixed pairings have been observed in NCSCs, e.g., superconducting gap nodes~\cite{YuanPhysRevLett2006,BonaldePhysRevLett2005,Xie2020,ShangPhysRevLett2020,PangPhysRevB2015}, multiband SC~\cite{ChenNJP2013,KuroiwaPhysRevLett2008,Harada200715390,ChenJPhysRevBYC}. On the other hand, \tcr{large upper critical field exceeding the Pauli limit is observed in some NCSCs}~\cite{BauerPhysRevLett2004,KimuraPhysRevLett2007,Carnicomeaar7969,KongPhysRevB2015, ArushiPhysRevB2020}.

Unconventional SC has been reported in different families of heavy fermion NCSCs, e.g., CePt$_3$Si~\cite{BauerPhysRevLett2004}, Ce$TX_3$ ($T$ = transition metal, $X$ = Si or Ge)~\cite{KimuraPhysRevLett2005,SugitaniJPSJ2006,SETTAI2007844,HONDA2010S543} and UIr~\cite{AkazawaJPCM2004}. In these compounds, the strong correlations or magnetic fluctuations can hinder the identification of the role played by the broken inversion symmetry in giving rise to unconventional superconducting properties, and as such non-magnetic weakly-correlated NCSCs have been investigated. For instance, while Li$_2$Pd$_3$B behaves as a fully-gapped two-band superconductor, an increase of ASOC (via Pt-for-Pd substitution) leads to Li$_2$Pt$_3$B being a nodal superconductor, indicating a dominant triplet component~\cite{YuanPhysRevLett2006,TakeyaPhysRevB2007,NishiyamaPhysRevLett2007}. In addition, a growing number of NCSCs have also been found to exhibit time-reversal symmetry (TRS) breaking (i.e., spontaneous magnetic fields) in their superconducting state, such as, CaPtAs~\cite{ShangPhysRevLett2020}, LaNiC$_2$~\cite{HillierPhysRevLett2009}, Zr$_3$Ir~\cite{ShangPhysRevBZr3Ir}, and \tcr{La$_7$$TM$$_3$ ($TM$ = Ni, Rh, Pd, Ir)}~\cite{Arushi73PhysRevB2021, SinghDPhysRevB2020, MayohPhysRevB2021, La7Ir3PhysRevLett2015}.
In some of these NCSCs, the ASOC is rather weak, and there is evidence for fully gapped superconductivity (SC) similar to $s$-wave superconductors, and the origin of the TRS breaking is generally not yet well understood.
	
Recently, the $\alpha$-Mn-type noncentrosymmetric Re-based superconductors have attracted considerable interest, mainly due to the observation of broken time reversal symmetry at the onset of SC~\cite{SinghPhysRevLett2014,SinghDPhysRevB2017,ShangPhysRevBRe24Ti5,ShangPhysRevLett2018}. On the other hand, in a few Re-free $\alpha$-Mn-type superconductors, e.g., Mg$_{10}$Ir$_{19}$B$_{16}$ and Nb$_{0.5}$Os$_{0.5}$~\cite{Acze2010,SinghNbOs}, TRS is preserved. Recent muon-spin relaxation ($\mu$SR) studies on Re$_{1-x}$Mo$_{x}$ alloys revealed that spontaneous magnetic fields below $T_c$ were observed only in elementary rhenium and in Re$_{0.88}$Mo$_{0.12}$~\cite{ShangPhysRevLett2018,shang2020re}, which both have centrosymmetric crystal structures. By contrast, TRS is preserved in the Re$_{1-x}$Mo$_{x}$ alloys for $x >$ 0.12, independent of their centro- or noncentrosymmetric crystal structures~\cite{shang2020re}. Moreover, TRS is preserved in both centro- or noncentrosymmetric rhenium-boron superconductors~\cite{2021PhysRevBShang}. All these results suggest that a noncentrosymmetric structure and thus the ASOC is not essential in realizing TRS breaking in Re-based superconductors, and its origins require further investigations.

The ZrNiAl-type compounds are another important family of noncentrosymmetric superconductors. (Zr,Hf)RuP~\cite{Barz3132} and ZrRu(As,Si)~\cite{MEISNER1983,Ichimin1999}, which were synthesized under high pressure, exhibit relatively high superconducting transition temperatures $T_c \sim$ 10\,K. Both ZrRuAs~\cite{DasPhysRevB2021} and LaPdIn~\cite{Su2021} are fully-gapped superconductors with preserved TRS. NbReSi also crystallizes in a ZrNiAl-type structure~\cite{osti1974}, but its physical properties are not yet well studied.

In this paper, we report a systematic study of the superconducting properties of non\-centro\-symmetric NbReSi by means of electrical-resistivity, magnetization, and heat-capacity measurements, as well as by electronic band-structure calculations. It is found that
NbReSi is a fully-gapped superconductor with a superconducting transition temperature $T_c$ = 6.5~K and exhibits a large upper critical field of $\mu_0H_{c2}$ = 12.6\,T. The electronic band-structure calculations suggest a sizable band splitting caused by the ASOC.

\section{Experimental details\label{sec:details}}\enlargethispage{8pt}

Polycrystalline NbReSi samples were prepared by arc melting stoichiometric amounts of Nb slugs (99.95$\%$, Alfa Aesar), Re powders (99.99$\%$, Alfa Aesar) and Si chunks (99.9999$\%$, Alfa Aesar) in high-purity argon atmosphere, with Ti metal used as oxygen getter. The samples were flipped and remelted several times to improve the homogeneity. The crystal structure and phase purity were checked by powder x-ray diffraction (XRD) measured on a Rigaku diffractometer with Cu K$\alpha$ radiation. The electrical-resistivity and heat-capacity measurements were performed on Quantum Design Physical Property Measurement System (PPMS) with a $^{3}$He cryostat. The magnetization measurements were carried out using Quantum Design Magnetic Property Measurement System (MPMS). The electronic band-structures were calculated by means of the density-functional theory (DFT) implemented in the Vienna \texttt{ab}-initio simulation package (VASP). The Perdew-Burke-Ernzerhoff (PBE) functional in the generalized gradient approximation was employed.

\begin{figure}
  	\includegraphics[angle=0,width=0.49\textwidth]{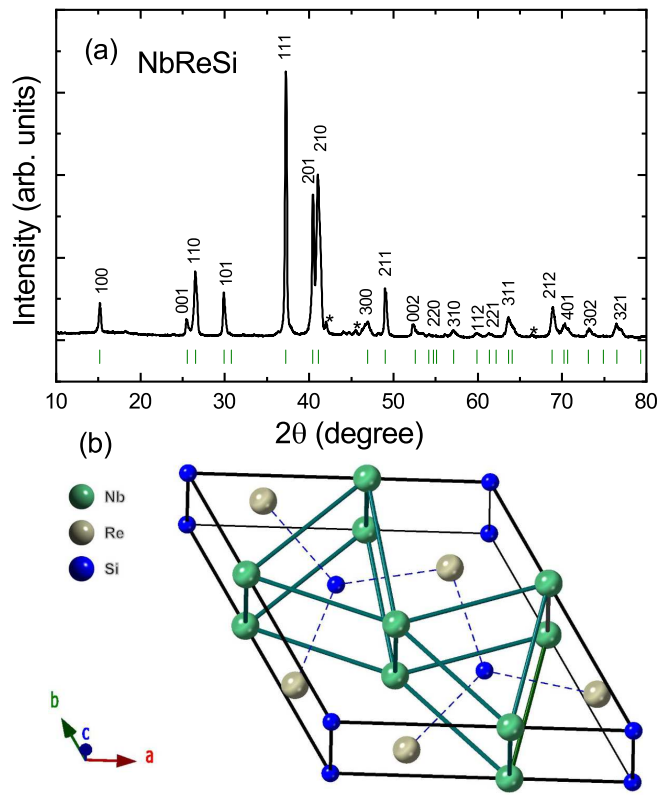}
  	\vspace{-12pt} \caption{\label{Figure1} (a) Powder x-ray diffraction pattern of NbReSi.
  	 The vertical bars are the calculated Bragg reflection positions for a ZrNiAl-type crystal structure.
  	 (b) Crystal structure of NbReSi. Nb, Re, and Si atoms are shown by green, grey, and blue spheres, respectively.}
\end{figure}

\begin{figure}
  	\includegraphics[angle=0,width=0.49\textwidth]{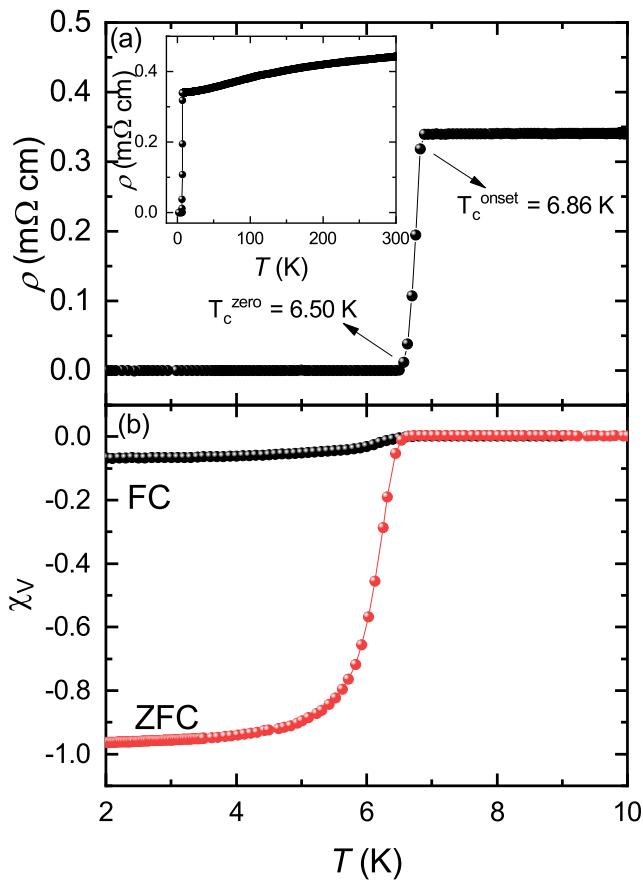}
  	\vspace{-12pt} \caption{\label{Figure2} (a) Temperature dependence of the electrical resistivity of NbReSi below 10\,K. The data between 2 and 300\,K is shown in the inset. (b) Temperature dependence of the magnetic susceptibility, measured in an applied field of 1 mT using both the ZFC and FC protocols. \tcr{The demagnetization factor is estimated to be about 0.2, considering the cuboid sample has the shape of $c/a$$\sim$1.4 and $c/b$$\sim$2.7 with field applied along $c$-direction \cite{AharoniJAP, OsbornPhysRev1945}.}}
\end{figure}

\begin{figure}
  	\includegraphics[angle=0,width=0.49\textwidth]{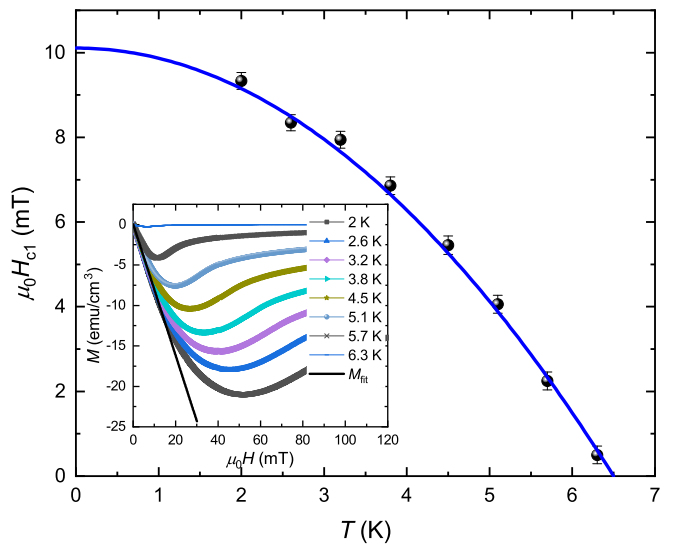}
  	\vspace{-12pt} \caption{\label{Figure3} \tcr{Lower critical fields $H_{c1}$ vs.\ temperature of NbReSi.
  			Solid line is a fit to $\mu_{0}H_{c1}(T) =\mu_{0}H_{c1}(0)[1-(T/T_{c})^2]$. The inset plots the
  		  field-dependent magnetization recorded at various temperatures. For each temperature, $H_{c1}$
  			was determined as the value where $M(H)$ starts deviating from
  			linearity (see black solid line).}}
\end{figure}

\begin{figure}
  	\includegraphics[angle=0,width=0.49\textwidth]{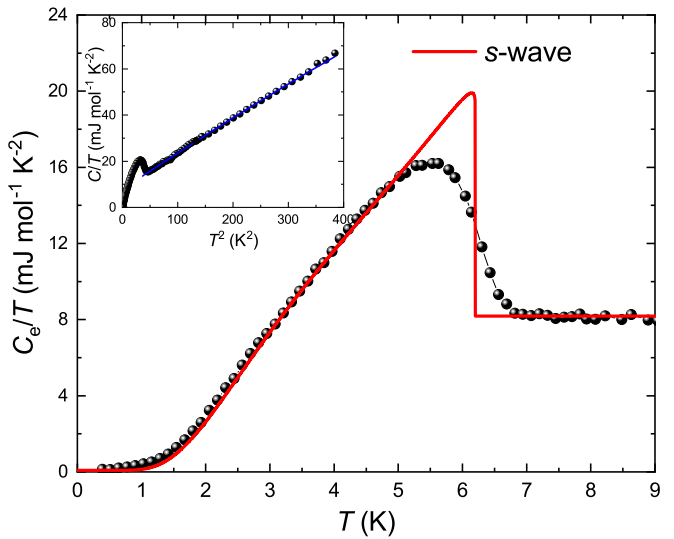}
  	\vspace{-12pt} \caption{\label{Figure4} Zero-field electronic specific heat
  		$C_e/T$ versus temperature for NbReSi. The red solid line through the data represents a fit to an $s$-wave model with a single gap. The inset shows the total specific heat $C/T$ versus $T^2$, where the blue line is a fit to  $C/T = \gamma_n + \beta T^2$.}
\end{figure}

\begin{figure}
  	\includegraphics[angle=0,width=0.45\textwidth]{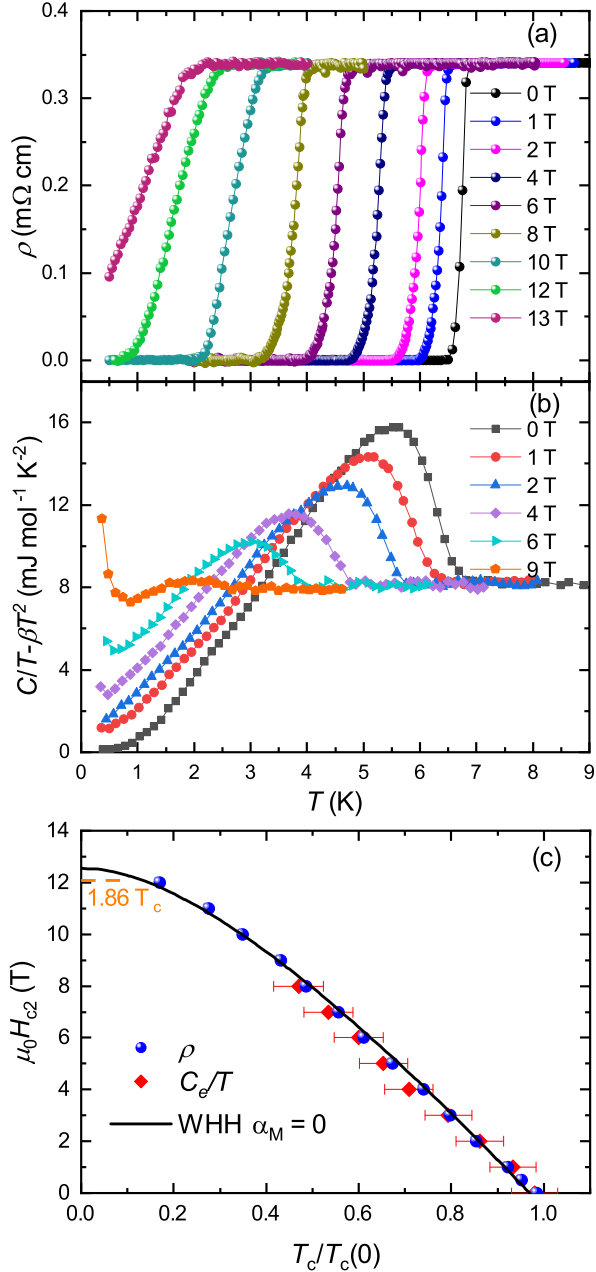}
  	\vspace{-12pt} \caption{\label{Figure5} Temperature dependence of the electrical resistivity (a) and specific heat (b) of NbReSi under various magnetic fields up to 13\,T. To better show the superconducting transitions, the phonon contribution $\beta$$T^2$ was subtracted from the specific heat. (c) The upper critical field $H_{c2}$ versus the reduced temperature $T_c$/$T_c(0)$ for NbReSi.
  		The solid line represents a fit to the  WHH model, while the dashed line marks the Pauli limit.}
\end{figure}

\begin{figure*}
  	\includegraphics[angle=0,width=0.98\textwidth]{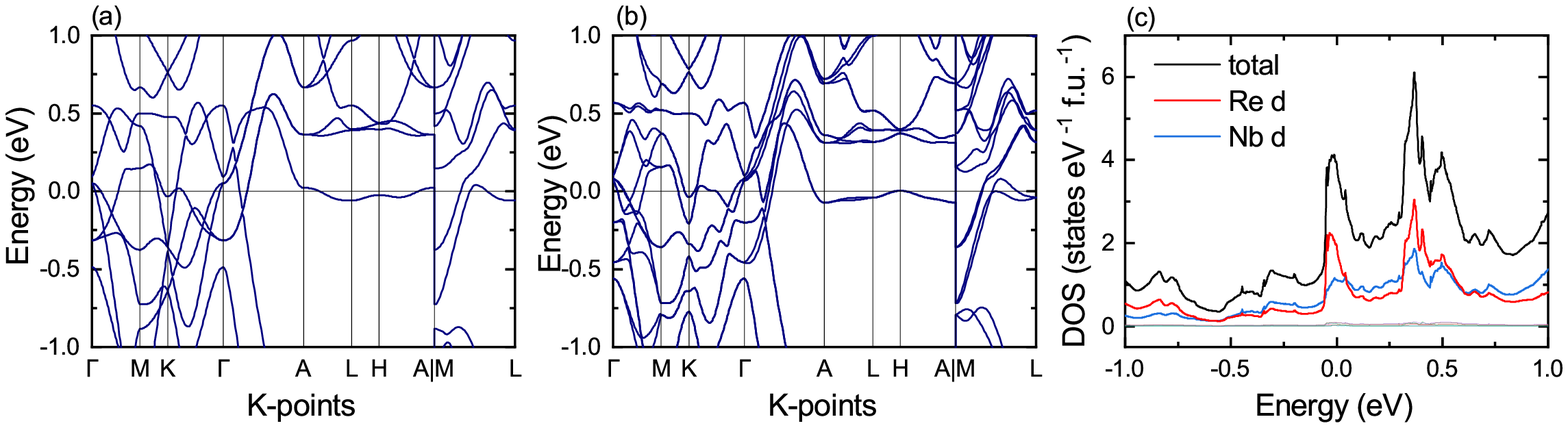}
  	\vspace{-12pt} \caption{\label{Figure6} Electronic band structure of NbReSi, calculated (a) without, and (b) with spin orbit coupling. (c) The total and partial (Re 5$d$ and Nb 4$d$ orbitals) densities of states near the Fermi level when SOC is considered. }
\end{figure*}

\section{ Results and Discussion\label{sec:results}}

Powder XRD patterns of NbReSi measured at room temperature are shown in Fig.~\ref{Figure1}(a). Almost all the reflections can be well indexed by a hexagonal ZrNiAl-type structure with space group $P\bar{6}2m$ (No.~189) (see vertical green bars). The peaks with rather weak intensity [marked by star symbols in Fig. \ref{Figure1}(a)] corresponds to additional phases, which could be due to binary Re-Si or Nb-Si phases that do not lead to extrinsic transitions. The determined lattice parameters $a$ = 6.7194(3)~\AA{} and $c$ = 3.4850(2)~\AA{} are consistent with the previous reports~\cite{osti1974}. The crystal structure of NbReSi is shown in Fig.~\ref{Figure1}(b), where the NbSi- and ReSi-layers are alternately stacked along the $c$-axis. In the unit cell, Nb and Re atoms occupy noncentrosymmetric 3$g$ and 3$f$ sites, while the two Si-sites (2$d$ and 1$a$) are centrosymmetric.

The temperature dependence of the electrical resistivity $\rho(T)$ [see inset of Fig.~\ref{Figure2}(a)], collected in zero magnetic field from 2 to 300\,K, reveals a metallic character of NbReSi. The main panel of Fig.~\ref{Figure2}(a) shows the enlarged plot of $\rho(T)$ below 10\,K.
There is a sharp superconducting transition below $T^\mathrm{onset}_c$ = 6.9\,K, reaching zero resistivity at $T^\mathrm{zero}_c$ = 6.5\,K, \tcr{which is slightly higher than the previous reported $T_c$ value (i.e., 5.1\,K)~\cite{rao1985structure}}.
The bulk SC of NbReSi was confirmed by the magnetic susceptibility measurements. The temperature dependence of the $dc$ magnetic susceptibility $\chi(T)$ of NbReSi, measured in a field of  1\,mT under field-cooled (FC) and zero-field-cooled (ZFC) processes, is shown in Fig.~\ref{Figure2}(b). A clear diamagnetic signal appears below the superconducting transition at $T_c$ = 6.5\,K. The large differences between the FC- and ZFC-susceptibilities are typical of type-II superconductors, where the magnetic flux is pinned once the material is cooled in an applied field. After accounting for the demagnetization factor, the superconducting shielding fraction is close to 100\%.

\tcr{To determine the lower critical field $H_\mathrm{c1}$, the field-dependent
	magnetization $M(H)$ of NbReSi was collected at various temperatures up to $T_c$ using a ZFC-protocol. Some representative $M(H)$ curves are shown in the inset of Fig.~\ref{Figure3}.
For each temperature, $H_\mathrm{c1}$ was determined as the value where $M(H)$ deviates from linearity (solid line in the inset of Fig.~\ref{Figure3}). 
Taking into account the demagnetization factor, the $H_\mathrm{c1}$ are summarized in the main panel of Fig.~\ref{Figure3} as a function of temperature, where the zero-temperature lower critical field $\mu_0H_\mathrm{c1}(0)$ = 10.1(1)\,mT is also determined.}
%

To study the SC of NbReSi, we also performed heat-capacity measurements. The jump in the specific heat at $T_c$ again demonstrates bulk SC in NbReSi (see Fig.~\ref{Figure4}).
In the normal state, the specific heat can be analyzed using $C(T)/T = \gamma_n$ + $\beta$$T^2$, where $\gamma_n$ is the normal-state electronic specific-heat coefficient and $\beta$$T^2$ represents the phonon contribution. As shown by the solid line in the inset of Fig.~\ref{Figure4}, $\gamma_n$ = 8.23(2)\,mJ\,mol$^{-1}$\,K$^{-2}$ and $\beta$ = 0.160(3)\,mJ\,mol$^{-1}$\,K$^{-4}$ were obtained for NbReSi. The Debye temperature $\Theta_D$ = 331\,K was calculated via $\Theta_D = (12\pi^4Rn/5\beta)^{1/3}$, where $R$ = 8.314\,J\,mol$^{-1}$\,K$^{-1}$ is the gas constant and $n = 3$ is the number of atoms per formula. The electron-phonon coupling constant $\lambda_{\mathrm{ep}}$
was estimated to be 0.66 using the McMillan formula~\cite{McMillanPhysRev1968}:
\begin{equation}\label{equation3}
\lambda_{\mathrm{ep}}=\frac{1.04+\mu^*\mathrm{ln}(\Theta_D/1.45T_c)}{(1-0.62\mu^*)\mathrm{ln}(\Theta_D/1.45T_c)-1.04},
\end{equation}
where the Coulomb pseudo-potential $\mu^*$ is fixed to a typical value of 0.13 for metallic materials. The density of states (DOS) at the Fermi level $N(E_F)$ can be estimated by $N(E_F)=3\gamma_n/\pi^2k_B^2$~\cite{kittel1996introduction}, which yields $N(E_F)$ = 3.47\,states\,eV$^{-1}$f.u.$^{-1}$.

After subtracting the phonon contribution, the zero-field
electronic specific heat $C_e/T$ of NbReSi is shown in Fig.~\ref{Figure4} as a function of temperature. The scaled specific jump at $T_c$, $\Delta$$C$/$\gamma_nT_c$ = 1.18, is smaller than the BCS value of 1.43. The reduced
specific-heat jump
at $T_c$ is typically attributed to multiband SC or gap anisotropy.
The contribution of SC to the electronic specific heat can be calculated via $C_e/T=\mathrm{d}S/\mathrm{d}T$, where $S$ is the entropy and can be written as~\cite{tinkham2004introduction}:
\begin{equation}\label{equation5}
S(T)=-\frac{3\gamma_n}{k_B\pi^3}\int_{0}^{2\pi}\int_{0}^{\infty}[(1-f)\mathrm{ln}(1-f)+f\mathrm{ln}f]\mathrm{d}\epsilon\mathrm{d}\phi.
\end{equation}
Here $f = (1 + e^{E/k_BT})^{-1}$ is the Fermi function and $E$ = $\sqrt{\epsilon^2 + \Delta^2(T)}$. The temperature-dependent superconducting energy gap
is approximated to $\Delta(T)$ = $\Delta(0)\tanh{1.82[1.018(T_c/T-1)]^{0.51}}$, where $\Delta(0)$ is the superconducting gap at 0\,K. The red solid line in Fig.~\ref{Figure4} represents a fit to the single-gap $s$-wave model. The derived gap value $\Delta(0)$ = 1.75\,$k_BT_c$ is close to the BCS value of 1.76\,$k_BT_c$, implying weak-coupling SC. It is noted that single-gap behaviors have been reported in other noncentrosymmetric Re-based superconductors, e.g., Re$_6$(Ti,Zr,Hf)~\cite{SinghDPhysRevB.972018,MayohPhysRevB2017,SinghDPhysRevB2016}, Re$_{24}$Ti$_5$~\cite{ShangPhysRevBRe24Ti5}, Re$_{0.82}$Nb$_{0.18}$~\cite{ShangPhysRevLett2018,ChenJPhysRevB2013}, Re$_3$W~\cite{BiswasPhysRevB2012}, and TaReSi~\cite{Sajilesh2021}.

To study the upper critical field $H_{c2}$ of NbReSi, the electrical resistivity and heat capacity were measured under various applied magnetic fields up to 13\,T.  As shown in Figs.~\ref{Figure5}(a) and (b), the superconducting transitions are gradually suppressed to lower temperatures and become broader with increases the magnetic field. When the magnetic field is larger than 4\,T, the specific heat starts to show an upturn at low temperatures, which becomes more prominent as increasing field and is likely due to the nuclear Schottky anomaly. With such an upturn anomaly, it is difficult to precisely determine the field dependence of the specific heat coefficient that might provide important insights into the gap symmetry. The upper critical field of NbReSi extracted from the electrical resistivity and specific heat are displayed in Fig.~\ref{Figure5}(c) as a function of the reduced temperature $T_c/T_c(0)$. Here $T_c$ is determined as the temperature where zero resistivity is reached, while for the specific heat, it is defined as the midpoint of specific-heat jump. The $H_{c2}$ values determined using different techniques are highly consistent.
It can be seen that the electrical resistivity reaches zero value before cooling to the lowest temperature in a field of 12\,T,
while in the 13\,T curve the transition can still been observed, indicating that the upper critical field $H_{c2}(0)$ at zero temperature
is slightly larger than the weak coupling Pauli limit value of 1.86$T_c$ = 12.1\,T. Moreover, the data agree well with the Werthamer-Helfand-
Hohenberg (WHH) model in the absence of paramagnetic limiting (i.e., with a Maki parameter $\alpha_M$ = 0)~\cite{WHHPhysRev1966}, yielding a zero-temperature value of $\mu_0H_{c2}(0)$ = 12.6~T.
The upper critical field larger than the weak-coupling Pauli limit has been reported in some other NCSCs, e.g., (Nb,Ta)Rh$_2$B$_2$~\cite{Carnicomeaar7969} and K$_2$Cr$_3$As$_3$~\cite{KongPhysRevB2015} and Y$_2$C$_3$~\cite{YUAN2011577}.
These results suggest the possibility that the effects of paramagnetic limiting may be reduced or absent in NbReSi superconductor. This can arise due to the Cooper pairs having a finite zero-temperature spin-susceptibility, which is a feature of both triplet SC and the admixture of spin-singlet and spin-triplet NCSCs under the influence of strong ASOC~\cite{Smidman2017}.
Alternatively, paramagnetic limiting fields larger than 1.86$T_c$ can also arise from a superconducting gap larger than the weak-coupling value, or deviations of the $g$-factor from the free electron value $g$ = 2~\cite{bauer2012non,ClogstonPhysRevLett1962}. While the former appears unlikely from our specific-heat analysis, confirmation of such a finite spin susceptibility requires studying the anisotropy of the upper critical fields, and utilizing nuclear magnetic resonance (NMR) to examine the Knight shift.

\tcr{The Ginzburg-Landau (GL) coherence length $\xi_{\mathrm{GL}}$ can be calculated from the upper critical field $\mu_0H_{c2}(0)$ using
\begin{equation}\label{equation8}
\mu_0H_{c2}(0)=\frac{\Phi_0}{2\pi\xi_{GL}^2},
\end{equation}
where $\Phi_0$ is the magnetic flux quantum, yielding $\xi_{\mathrm{GL}}$ = 5.11\,nm for NbReSi. The magnetic penetration depth $\lambda_{\mathrm{GL}}$ = 243.8~nm can be obtained via:
\begin{equation}\label{equation9}
\mu_0H_{c1}(0)=\frac{\Phi_0}{4\pi\lambda_{\mathrm{GL}}^2}\mathrm{ln}\frac{\lambda_\mathrm{GL}}{\xi_\mathrm{GL}}.
\end{equation}
The GL parameter $\kappa_{\mathrm{GL}}$ = $\lambda_{\mathrm{GL}}/\xi_{\mathrm{GL}}$ = 49.3 is much larger than the threshold value of 1/$\sqrt{2}$, indicating that NbReSi is a strongly type-II superconductor.
The thermodynamic critical field $\mu_0H_c(0)$ can be estimated from
\begin{equation}\label{equation10}
	\mu_0H_{c1}(0)\mu_0H_{c2}(0) = \mu_0H_{c}(0)^2\mathrm{ln}\kappa_{\mathrm{GL}},
\end{equation}
yielding a value of 181~mT for NbReSi.
}

To get more insight into the underlying electronic properties of NbReSi, we also performed the electronic band-structure calculations based on DFT, with and without considering the spin-orbit coupling.
As shown in Fig.~\ref{Figure6}, several bands crossing the Fermi level can be identified. By taking into account the SOC, the band splitting $E_{\mathrm{ASOC}}$ near the Fermi level is estimated to be 180\,meV. Its energy scale to the
superconducting energy gap,
i.e., $E_{\mathrm{ASOC}}/k_BT_c$ $\sim$ 350, is relatively large compared with many other NCSC~\cite{Smidman2017}.
Such a large band splitting suggests an significant effect of ASOC. Relatively more obvious band splitting along $c$-axis
related to the high symmetry lines $\Gamma$-A and M-L suggest a possible anisotropy in NbReSi, which remains verified on single crystal measurements.
Figure~\ref{Figure6}(c) shows the total and partial density of states (DOS) for NbReSi within the energy scale of -1 to 1\,eV.
The calculated DOS at the Fermi level $N(E_F)$ $\sim$ 3.94\,states~eV$^{-1}$\,f.u.$^{-1}$ is compatible with the value 3.47~states\,eV$^{-1}$\,f.u.$^{-1}$ determined from the normal-state electronic specific-heat coefficient.
The contributions to $N(E_F)$ mainly arise from Re-5$d$ and Nb-4$d$ orbits, while the contributions from other orbits or orbits of Si atoms are negligible.
Both Re 5$d$ and Nb 4$d$ orbits exhibit relatively large SOC, which results in
the large band splitting near the Fermi level.

In noncentrosymmetric superconductors, the ASOC in principle can lift the degeneracy of the electronic bands and thus, the admixture of spin-singlet and spin-triplet pairing states is allowed. For NbReSi, the specific-heat results suggest a fully-gapped superconducting state, more consistent with spin-singlet pairing. To further understand the superconducting pairings and the effect of ASOC in NbReSi, the $\mu$SR or NMR studies will be very helpful.
For comparison, the superconducting- and normal-state properties of NbReSi, as well as the recently reported NCSC TaReSi are summarized in Table~\ref{table:table1}.
Different from NbReSi, TaReSi crystallizes in an orthorhombic TiFeSi-type structure ($Ima$2, No.~46)(see Table.~\ref{table:table1})~\cite{Sajilesh2021} . Though TaReSi exhibits a comparable $T_c$ value to
NbReSi and fully-gapped SC, its upper critical field is much smaller, well below the Pauli limit, i.e., $H_{c2}/H_p$ $\sim$ 0.68. Whether such distinct upper critical fields
in these two compounds are related to the strength of ASOC or other factors requires further investigations.

The large $H_{c2}$ that is comparable to the Pauli limit seems to be a general feature in Re-based superconducting
alloys, including the $\alpha$-Mn type materials and NbReSi,
and centrosymmetric Re$_3$W, in spite of the preserved or broken TRS in their superconducting states~\cite{Smidman2017, BiswasPhysRevB2011, ShangFIP2021}. The origin of a large $H_{c2}$ is not yet fully understood.
It is also possible that the $H_{c2}$ can be enhanced due to the presence of
disorder and nonmagnetic impurities~\cite{GurevichPhysRevB2003}. The upper critical field exceeds the weak coupling Pauli limit in NbReSi,
but the determination of whether paramagnetic limiting is weakened or absent in this compound requires further studies on high-quality single crystal.

In addition, the observation of TRS breaking in some
Re-based superconductors suggests an important role played by Re.  While the previous $\mu$SR studies on the Re-free ZrNiAl-type ZrRuAs and LaPdIn suggest a preserved TRS in their superconducting states~\cite{DasPhysRevB2021,Su2021}.
Therefore,
it is of particular interest to determine whether the TRS is broken in the superconducting state of NbReSi, which might further shed light on the origin of TRS breaking in the Re-based superconductor.

\begin{table}[tb]
	\caption{Superconducting and normal-state properties of NbReSi and TaReSi. Data of TaReSi were taken from Ref.~\onlinecite{Sajilesh2021}}.
	\label{table:table1}
	\begin{ruledtabular}
		\begin{tabular}{c c c c }
			Parameters &Unit  &NbReSi  &TaReSi  \\
			\hline\\[-2ex]
			{Space group}  &{-}    &$P\bar{6}2m$     &$Ima$2 \\
			{$T_c$}  &K    &6.5     &5.32 \\
           \tcr{$\mu_0H_{c1}(0)$} &mT   &10.1     &6.27 \\
			{$\mu_0H_{c2}(0)$}  &T  &12.6  &6.6  \\
		    {$\mu_0H_{c}(0)$}   &mT  &181  &114 \\
			{$\mu_0H_P(0)$} &T  &12.1  &9.73  \\
         \tcr{$\lambda_{\mathrm{GL}}(0)$}     &\AA  &2519  &3373  \\
		 \tcr{$\xi_{\mathrm{GL}}(0)$}     &\AA  &51     &137                        \\
			$\theta_\mathrm{D}$      &K   &331   &338   \\
			$N(E_\mathrm{F})$ &{states eV$^{-1}$ f.u.$^{-1}$}  &3.94  &2.28   \\
			$\Delta$$C_{\mathrm{el}}$/$\gamma_n$$T_c$ &{-}  &1.18  &1.07  \\
			$\Delta(0)$/$k_\mathrm{B}T_c$ &{-}  &1.75  &1.4  \\
		\end{tabular}
	\end{ruledtabular}
\end{table}

\section{Summary\label{sec:summary}}

To summarize, the noncentrosymmetric NbReSi superconductor were synthesized and investigated
by means of the electrical-resistivity, magnetization, and heat-capacity measurements, as well as via electronic band structure calculations. We found that NbReSi shows bulk superconductivity at $T_c$ = 6.5\,K.
Its upper critical field $\mu_0H_{c2}(0)$ = 12.6\,T is comparable to the weak coupling Pauli limit.
The low-temperature zero-field electronic specific heat data suggest nodeless SC, with a gap value close to the BCS theoretical value.
The specific-heat discontinuity and the electron-phonon coupling constant $\lambda_\mathrm{ep}$ suggest weak-coupling SC in NbReSi. Electronic band structure calculations reveal a relatively large band splitting near the Fermi level due to the presence of a strong ASOC. These results suggest that NbReSi represents a new candidate material to study the broken TRS in the superconducting state of weakly correlated NCSCs.

\section{Acknowledgments}

This work was supported by the Key R\&D Program of Zhejiang Province, China (2021C01002), the National Natural Science Foundation of China (No. 11874320, No. 12034017, and No. 11974306), and the National Key R\&D Program of China (No. 2017YFA0303100). T. Shang acknowledge the support from the Natural Science Foundation of Shanghai (Grant No. 21ZR1420500 and No. 21JC1402300).

\bibliography{NbReSiref}

\end{document}